\documentclass[aps,prl,superscriptaddress,amsmath,amssymb,
showpacs,twocolumn,floatfix]{revtex4}
\usepackage{graphicx,bm}  
\usepackage{subfigure}
 \usepackage{epsfig} 
\begin{document}
\def\bar{\begin{eqnarray}}
\def\ear{\end{eqnarray}}
\def\beq{\begin{equation}}
\def\eeq{\end{equation}}
\newcommand{\degrees}{\ensuremath{^{\circ}}}
\newcommand{\bsigma}{\mbox{\boldmath$\sigma$}}
\newcommand{\bnabla}{\mbox{\boldmath$\nabla$}}
\title{Odd Triplet Pairing in clean Superconductor/Ferromagnet
heterostructures}
\author{Klaus Halterman}
\email{klaus.halterman@navy.mil}
\affiliation{Physics and Computational Sciences, Research and Engineering Sciences Department, Naval Air Warfare Center,
China Lake, California 93555}
\author{Paul H. Barsic}
\email{barsic@physics.umn.edu}
\author{Oriol T. Valls}
\email{otvalls@umn.edu}
\altaffiliation{Also at Minnesota Supercomputer Institute, University of Minnesota,
Minneapolis, Minnesota 55455}
\affiliation{School of Physics and Astronomy, University of Minnesota, 
Minneapolis, Minnesota 55455}
\date{\today}


\begin{abstract}

We study triplet pairing correlations 
in clean  Ferromagnet (F)/Superconductor (S)
nanojunctions, via fully self consistent solution of
the Bogoliubov-de Gennes equations.
We consider  FSF  trilayers, with S being an 
s-wave superconductor, and an arbitrary angle
$\alpha$ between the magnetizations of the two F layers. We find that
contrary to some previous expectations,  
triplet correlations, odd in time, are  induced in both 
the S and F layers in the clean limit.
We investigate their behavior as a function of time, position, and
$\alpha$. 
The triplet amplitudes are largest at 
times on the order of the inverse
``Debye'' frequency, and at that time scale they are long ranged
in both S and F. The  
zero temperature 
condensation energy 
is found to be lowest  when
the magnetizations are antiparallel.

\end{abstract}
\maketitle


The proximity effects in superconductor/ferromagnet (SF) heterostructures lead
to the coexistence of ferromagnetic and superconducting ordering
and to novel transport phenomena\cite{buzdinR,bergR}. Interesting
effects that arise from the  interplay
between these orderings have potential
technological applications in fields such as spintronics\cite{zutic}.  For
example, 
the relative orientation of 
the magnetizations in the F layers in FSF trilayers can
have a strong influence on the 
conductivity\cite{gu,moraruL,bell,visani,hv72}, making them good
spin valve candidates.  Such trilayers were first proposed\cite{dg66} for 
insulating F layers and later for metallic\cite{tagirov,buzdin99} ones.

This 
interplay also results in fundamental new physics. 
An outstanding example is the
existence of ``odd'' triplet
superconductivity.  This is an 
{\it s-wave} pairing triplet state that is
even in momentum, and therefore
not destroyed by nonmagnetic impurities, but with the triplet correlations 
being odd in frequency,
so that the equal time triplet amplitudes
vanish as required by the Pauli principle.
This exotic pairing state 
with total spin one 
was 
proposed long ago 
\cite{berez} as a possible  state in  
superfluid $^3{\rm He}$. Although this type of pairing does not occur there,  
it is possible in certain FSF systems\cite{buzdinR,bergR,berg86,berg68} 
with ordinary
singlet pairing in S. 
This arrangement can induce, via proximity effects,
triplet correlations with $m=0$ and $m=\pm1$ projections
of the total spin.
If the magnetization orientations in both F layers
are unidirectional and along the quantization axis, symmetry arguments show 
that only the $m=0$ projection along that axis can exist. 

Odd triplet
pairing in F/S structures has been studied in the dirty limit through 
linearized Usadel-type quasiclassical 
equations \cite{bergR,berg86,berg68,eschrig2}. 
In this case, 
it was found that 
$m=0$ triplet pairs always exist. They are
suppressed in F over short length scales, just as the singlet pairs.
The $m=\pm1$ components, for which the exchange field is not pair-breaking,
can be long ranged, and  
were found to exist for nonhomogeneous magnetization. 
For FSF trilayers\cite{bergR,fominov,eschrig},
the 
quasiclassical methods predict that the  structure 
contains a superposition of all three 
spin triplet projections
except when the magnetizations of the F layers are collinear, 
in which case the $m=\pm1$ components along the magnetization
axis vanish.
It is noted in Ref.~\cite{buzdinR}  
that the existence
of such effects in the clean limit has not been established 
and may be doubted.
This we remedy
in the present work, where we establish that, contrary to the 
doubts voiced there, induced, long-ranged, odd triplet pairing does occur in 
clean FSF structures.

\begin{figure}
\centering
\includegraphics[width=3in]{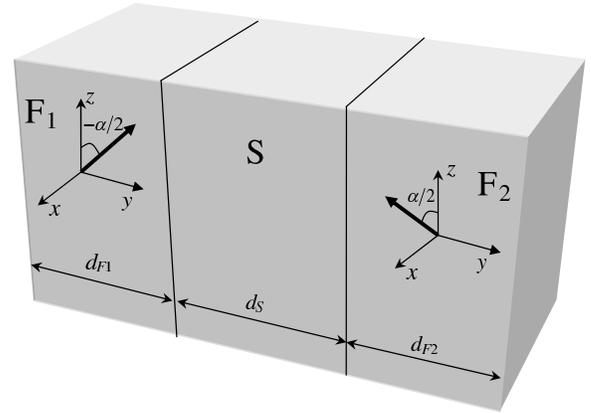}
\caption{Schematic of FSF junction. The left ferromagnetic layer 
$\rm F_1$ has a  magnetization oriented at an angle $-\alpha/2$ in the $x-z$ plane, while
the other ferromagnet, $\rm F_2$, has a magnetization orientation at an
angle $\alpha/2$  in the $x-z$ plane.}
\label{fig1} 
\end{figure}

Experimental results that may argue for the existence of
long range triplet pairing of superconductors through a ferromagnet 
have been obtained in
superlattices\cite{pena} with ferromagnetic spacers, and in  two
superconductors coupling through a single ferromagnet\cite{nelson,keizer}. 
Measurements\cite{nelson} on a SQUID, in which a phase change of $\pi$ 
in the order parameter  is found after inversion, indicate an odd-parity state.
Very recently,  a Josephson current through  a strong  
ferromagnet was observed, indicating the existence of a spin triplet state\cite{keizer}
induced by 
${\rm NbTiN}$, an s-wave superconductor.

In this paper, we study the induced odd triplet 
superconductivity in FSF trilayers in the
clean limit through a fully self-consistent 
solution of the microscopic Bogoliubov-de Gennes (BdG) equations.  
We consider  arbitrary 
relative orientation of the magnetic moments in the two F 
layers.   We find that there are indeed induced
odd triplet correlations which  
can include both
$m=0$ and $m=\pm1$ projections.  We directly
study their time dependence and we find that
they are largest for times of order of the inverse cutoff ``Debye'' 
frequency.  The correlations are, at these time
scales, 
long ranged in both the S and F regions.
We also find that the condensation energy depends on the relative
orientation of the F layers, being a minimum when they are antiparallel.

To find the triplet correlations arising
from the nontrivial spin structure in
our FSF system,
we use the BdG equations with
the BCS Hamiltonian, ${\cal H}_{\rm eff}$: 
\begin{widetext}
\begin{align}
{\cal H}_{\rm eff} &=\int d^3 r\Bigl\lbrace \sum_\delta 
\psi^\dagger_\delta({\bf r})  
\left[-\frac{\bnabla^2}{2m^*}-E_F \right]\psi_\delta({\bf r}) 
+ \frac{1}{2} [\sum_{\delta,\beta}  (i\sigma_y)_{\delta \beta} 
\Delta({\bf r}) \psi^\dagger_\delta({\bf r}) \psi^\dagger_\beta({\bf r})+ 
\rm{h.c.}] \nonumber 
 -\sum_{\delta,\beta} \psi^\dagger_\delta({\bf r})({\bf h} \cdot 
\bsigma)_{\delta \beta}\,\psi_\beta({\bf r}) \Bigr\rbrace,
\end{align}
\end{widetext}
where 
$\Delta({\bf r})$ is the pair potential, to be determined self-consistently, 
$\psi^\dagger_\delta,\psi_\delta$ are the creation and annihilation operators
with spin $\delta$, $E_F$ is the Fermi energy,  
and ${\boldsymbol \sigma}$ are the Pauli matrices.
We describe the magnetism of
the F layers by an effective exchange field ${\bf h(r)}$ that
vanishes in the S layer. We will consider the geometry 
depicted in Fig.~\ref{fig1}, 
with  the  $y$ axis normal
to the layers and ${\bf h(r)}$ in the $x-z$ plane (which is infinite in extent)  
forming an angle $\pm \alpha/2$ with the $z$
axis in each F layer. 

Next, we expand the field operators in terms of a Bogoliubov transformation
which we write as:
\begin{equation}
\label{bv}
\psi_{\delta}({\bf r})=\sum_n \left(u_{n\delta}({\bf r})\gamma_n +\eta_\delta v_{n\delta}({\bf r})\gamma_n^\dagger\right), 
\end{equation}
where $\eta_\delta\equiv 1(-1)$ for spin down (up), 
$u_{n\delta}$ and $v_{n\delta}$ are the quasiparticle and quasihole
amplitudes.
This transformation diagonalizes ${\cal H}_{\rm eff}$:
$[{\cal H}_{\rm eff},\gamma_n]=-\epsilon_n\gamma_n$, 
$[{\cal H}_{\rm eff},\gamma^\dagger_n]=\epsilon_n\gamma^\dagger_n$.
By taking the commutator $[\psi_{\delta}({\bf r}),{\cal H}_{\rm eff}]$, and with 
${\bf h(r)}$ in the $x-z$ plane as explained above, we have the following:
\begin{subequations}
\label{comm}
\begin{align}
[\psi_{\uparrow}({\bf r}),{\cal H}_{\rm eff}]&=({\cal H}_{\rm e}-h_z)\psi_{\uparrow}({\bf r})-h_x\psi_{\downarrow}({\bf r})+ 
\Delta({\bf r})\psi^\dagger_{\downarrow}({\bf r}), \\
[\psi_{\downarrow}({\bf r}),{\cal H}_{\rm eff}]&=({\cal H}_{\rm e}+h_z)\psi_{\downarrow}({\bf r})-h_x\psi_{\uparrow}({\bf r})- 
\Delta({\bf r})\psi^\dagger_{\uparrow}({\bf r}).
\end{align}
\end{subequations}
Inserting (\ref{bv}) into (\ref{comm}) 
and introducing a set ${\bm \rho}$ of Pauli-like matrices in particle-hole
space,  
yields the spin-dependent BdG equations:
\begin{equation}
\left[
\rho_z\otimes\left({\cal H}_0 \hat{\bf 1} -h_z\sigma_z\right)+\left(\Delta(y)\rho_x
-h_x \hat{\bf 1}\right)\otimes \sigma_x \right]{\Phi}_n=\epsilon_n{\Phi}_n,
\label{bogo}
\end{equation}
where 
${\Phi}_n
\equiv(u_{n\uparrow}(y),u_{n\downarrow}(y),v_{n\uparrow}(y),v_{n\downarrow}(y))^{\rm T}$
and ${\cal H}_0\equiv-\partial^{2}_{y}/(2m)+\varepsilon_{\perp} 
-E_F$.
Here $\varepsilon_{\perp}$ is  the transverse kinetic 
energy and a factor of $e^{i{\bf k_\perp \cdot r}}$ has been suppressed. 
In deriving Eq.~(\ref{bogo}) care has been taken to consistently use the
phase conventions in Eq.~(\ref{bv}). 
To find the quasiparticle amplitudes 
along a different quantization axis in the $x-z$ plane,
one performs a spin rotation: 
$\Phi_n\rightarrow\widehat{U}(\alpha^\prime){\Phi}_n$,
where $\widehat{U}(\alpha^\prime)=
\cos(\alpha^\prime/2) \hat{\bf 1}\otimes\hat{{\bf 1}}-i\sin(\alpha^\prime/2)\rho_z\otimes\sigma_y$. 

When 
the magnetizations of the F layers are collinear,
one can take $h_x=0$.
For the  general case  
shown in Fig.~\ref{fig1} one has
in the $\rm F_1$ layer,
$h_x=h_0 \sin(-\alpha/2)$ and $h_z=h_0 \cos(-\alpha/2)$, where
$h_0$ is the magnitude of ${\bf h}$, while in $\rm F_2$,
$h_x=h_0 \sin(\alpha/2)$, and $h_z=h_0 \cos(\alpha/2)$.  
With an appropriate choice of basis, 
Eqs.~(\ref{bogo}) are cast into a matrix eigenvalue system that is
solved iteratively
with the
self consistency condition, $\Delta(y)=g(y) f_3$ 
($f_3 = \frac{1}{2}\left[\langle \psi_{\uparrow}({\bf r}) \psi_{\downarrow} ({\bf r})\rangle-
\langle \psi_{\downarrow}({\bf r}) \psi_{\uparrow} ({\bf r})\rangle\right]$).
In the F layers we have $g(y)=0$,
while in S,
$g(y)=g$, $g$ being the usual BCS 
{\it singlet} coupling constant there. 
Through Eqs.~(\ref{bv}), the self-consistency condition becomes a sum 
over states
restricted by the factor $g$
to  within  
$\omega_D$ from the Fermi surface. Iteration is performed until
self-consistency is reached. 
The numerical  process is  the same that was used
in previous work\cite{hv69,hv70}, with now the $h_x$ term requiring larger
four-component matrices to be diagonalized.

We now define the following time dependent
{\it triplet}  amplitude functions in terms of the field operators,
\begin{subequations}
\label{pa}
\begin{align}
\tilde{f}_0({\bf r},t) =& \frac{1}{2}\left[
\langle \psi_{\uparrow}({\bf r},t) \psi_{\downarrow} ({\bf r},0)\rangle+
\langle \psi_{\downarrow}({\bf r},t) \psi_{\uparrow} ({\bf r},0)\rangle\right],\\
\tilde{f}_1({\bf r},t) =& \frac{1}{2}\left[
\langle \psi_{\uparrow}({\bf r},t) \psi_{\uparrow} ({\bf r},0)\rangle-
\langle \psi_{\downarrow}({\bf r},t) \psi_{\downarrow} ({\bf r},0)\rangle 
\right],
\end{align}
\end{subequations}
which, as required by the Pauli principle
for these $s$-wave amplitudes, vanish at $t=0$, as we shall verify.
Making use of  Eq.~(\ref{bv})  and  the 
commutators, 
one can derive and formally integrate the 
Heisenberg equation of the motion for the operators and obtain:
\begin{subequations}
\label{pa3}
\begin{align}
\tilde{f}_0(y,t) =& \frac{1}{2}\sum_n[u_{n\uparrow}(y) v_{n\downarrow}(y)-u_{n\downarrow}(y) 
v_{n \uparrow}(y)]
\zeta_n(t), \label{f0} \\
\tilde{f}_1(y,t) =& -\frac{1}{2}\sum_n [u_{n\uparrow}(y) v_{n\uparrow}(y)+u_{n\downarrow}(y) v_{n\downarrow}(y)] 
\zeta_n(t),
\end{align}
\end{subequations}
where $\zeta_n(t)\equiv \cos(\epsilon_n t)-i\sin(\epsilon_n t)\tanh(\epsilon_n/2T)$.

\begin{figure}
\centering
\includegraphics[width=3in]{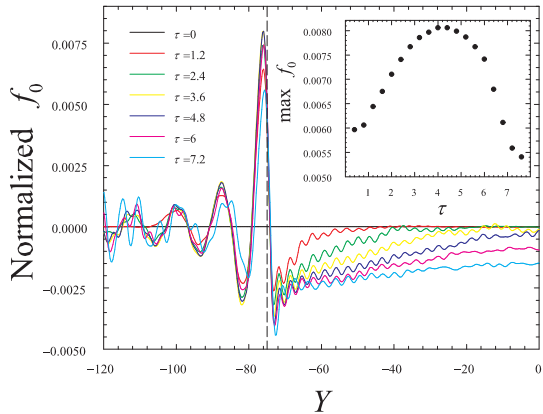}
\caption{(Color online) The real part, $f_0$, of the triplet amplitude $\tilde{f}_0$,  
for a FSF trilayer
at 7 different times. We normalize $f_0$ by  the singlet bulk pair amplitude, 
$\Delta_0/g$. The coordinate $y$ is
scaled by the Fermi wavevector, 
$Y\equiv k_F y$, and time by the Debye frequency,
$\tau\equiv\omega_D t$. 
At $\tau=0$, $f_0\equiv 0$ as required by the Pauli principle.
The interface is marked by the vertical dashed line, with an F region  to
the left and the S to the right.
Half of the S region and  part of the left F layer are shown.
The inset shows the maximum
value of $f_0$ versus $\tau$.} 
\label{fig2} 
\end{figure}

The  amplitudes in Eqs.~(\ref{pa3})
contain all information on the space and time dependence of induced triplet 
correlations
throughout the FSF structure.
The summations in Eqs.~(\ref{pa3})
are over the entire self-consistent spectrum,
ensuring that $f_0$ and $f_1$ vanish identically at $t=0$ 
and thus obey the exclusion principle. 
Using a non self consistent $\Delta(y)$ leads to violations of this
condition,
particularly near the interface where proximity effects are most pronounced. 
Geometrically, the indirect coupling between magnets is stronger 
with fairly thin S layers
and relatively thick F layers.
We thus have chosen
$d_S=(3/2)\xi_0$ and $d_{F1}=d_{F2} = \xi_0$, with 
the BCS correlation length $\xi_0=100 k_F^{-1}$. 
We consider the low $T$ limit and take
$\omega_D=0.04 E_F$.
The magnetic exchange  is parametrized via $I\equiv h_0/E_F$.
Results shown are for $I=0.5$ (unless otherwise noted)  
and the magnetization orientation angle, $\alpha$, is swept over the
range $0\leq\alpha\leq\pi$. 
No triplet amplitudes arise in the absence of magnetism ($I=0$).

For the time scales considered here, the imaginary 
parts of $\tilde{f}_0(y,t)$ and $\tilde{f}_1(y,t)$ 
at $t\neq 0$ are considerably smaller than their real
parts,
and thus we focus 
on the latter, which we denote  by $f_0(y,t)$ and $f_1(y,t)$. 
In Fig.~\ref{fig2}, the spatial dependence of  $f_0$ 
is shown for parallel magnetization directions ($\alpha=0$)
at several times $\tau\equiv\omega_D t$.
The spatial range shown includes part of the $F_1$ layer (to the left of the
dashed line)
and half of the S layer (to the right).
At finite $\tau$, the maximum  occurs in the ferromagnet 
close to the interface,
after which $f_0$ undergoes damped oscillations with the usual spatial 
length scale $\xi_f\approx (k_{F\uparrow}-k_{F\downarrow})^{-1}\approx k_F^{-1}/I$.
The height of the main peak first increases with time, but drops off
after a characteristic time, $\tau_c\approx 4$, as
seen in the inset, which
depicts the  maximum value of $f_0$ as a function of  $\tau$. 
As $\tau$ increases beyond $\tau_c$, the modulating $f_0$ in F develops
more complicated atomic scale interference patterns and becomes
considerably longer ranged.
In S, we see immediately
that $f_0$ is also larger near the interface.
Since the triplet amplitudes vanish at $\tau=0$, short time scales
exhibit correspondingly short triplet penetration. 
The figure shows, however, that the value of  $f_0$ in S is substantial
for $\tau\gtrsim \tau_c$, 
extending over  length scales on the order of $\xi_0$ without
appreciable decay.
In contrast, the usual singlet correlations were found to monotonically 
drop off from their $\tau=0$ value
over $\tau$ scales of order unity. 

\begin{figure}
\centering
\includegraphics[width=3in]{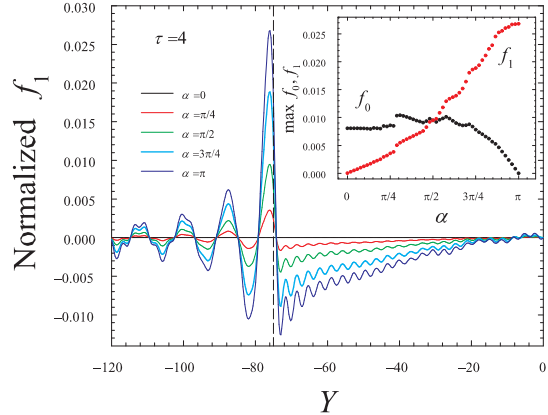}
\caption{(Color online) Spatial and angular dependence of  $f_1$, 
at $\tau=4 \approx \tau_c$ and several $\alpha$. Normalizations and ranges
are as in Fig.~\ref{fig2}. Inset:
maxima of $f_0$ and $f_1$ in ${\rm F}_1$ versus $\alpha$. 
}
\label{fig3} 
\end{figure}

In the main plot of Fig.~\ref{fig3} we examine the spatial dependence 
of the real part of the $m=\pm1$ triplet 
amplitude, $f_1$. 
Normalizations and spatial ranges are
as in Fig.~\ref{fig2} 
but now the
time is fixed at $\tau=4\approx\tau_c$, and five equally spaced magnetization 
orientations are considered.
At $\alpha=0$, $f_1$ vanishes identically at all $\tau$, as expected. 
For nonzero $\alpha$, correlations in
all triplet channels are present.
As was found for $f_0$, the plot clearly shows  
that $f_1$ is largest near the interface, in the ${\rm F}$ region. 
Our geometry and conventions imply (see Fig.~\ref{fig1}) 
that the magnetization 
has opposite  $x$-components in the  ${\rm F}_1$ and ${\rm F}_2$ regions.
The $f_1$ triplet pair amplitude profile is thus antisymmetric about the 
origin, in contrast to  
the symmetric
$f_0$, implying the existence of one node in the superconductor.
Nevertheless, the penetration  of the $f_1$ correlations in S
can  be long ranged. 
We find that $f_1$ and $f_0$
oscillate in phase and with the same wavelength, regardless of $\alpha$.
The inset illustrates the maximum attained values of $f_0$ and $f_1$ in ${\rm F}_1$ as 
$\alpha$  varies. 
It shows that
for a broad range of $\alpha$, $\alpha\lesssim 3\pi/4$, the
maximum of $f_0$ varies relatively 
little, after which it drops off rapidly to zero
at $\alpha=\pi$. 
This is to be expected as the anti-parallel orientation corresponds to the case
in which the magnetization is in the $x$ direction, which is perpendicular to the
axis of quantization (see Fig.~\ref{fig1}).
The rise in the maximum of $f_1$ is monotonic, cresting at $\alpha=\pi$,  
consistent with the main plot. At this angle
the triplet correlations  extend considerably 
into the superconductor. 
At $\alpha=\pi/2$ the maxima coincide since
the two triplet components 
are then identical throughout the whole space because
the magnetization vectors have equal projections on the $x$
and $z$ axes. At $\alpha=\pi$ both magnetizations are
{\it normal} to the axis of quantization $z$ (see
Fig.~\ref{fig1}).
By making use of the rotation matrix $\widehat{U}$ (see below Eq.~\ref{bogo})
one can verify that the $m=\pm 1$ components with respect to the axis 
$x$ along the magnetizations
are zero.  

\begin{figure}
\centering
\includegraphics[width=3in]{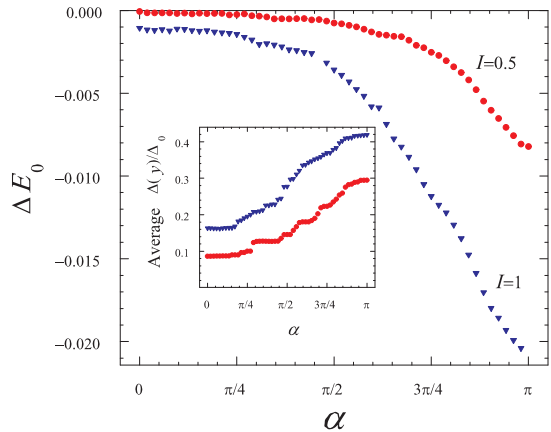}
\caption{(Color online) The $T=0$ condensation energy, $\Delta E_0$, 
normalized by $N(0)\Delta_0^2$ ($N(0)$ is the usual density of states), 
vs. the  angle $\alpha$ for two values of $I$. 
When the two magnetizations
are antiparallel ($\alpha=\pi$) $\Delta E_0$
is  lowest. The inset shows the ordinary 
(singlet)
pair potential averaged over the S region, normalized to the bulk $\Delta_0$.}
\label{fig4} 
\end{figure}

We next consider the condensation energy, $\Delta E_0$,
calculated by subtracting the zero temperature superconducting and
normal state free energies.
The calculation uses the self
consistent spectra and $\Delta(y)$, and
methods explained elsewhere \cite{hv70,kos}.
In the main plot of Fig.~\ref{fig4}, we show 
$\Delta E_0$ (normalized at twice its bulk S value)
at two different values of $I$.
The condensation energy results clearly demonstrate that the antiparallel 
state ($\alpha=\pi$) is 
in general the lowest energy ground state. 
These results
are consistent with previous studies\cite{hv72} of FSF structures with 
parallel and antiparallel magnetizations. 
The inset contains the magnitude of the spatially averaged pair potential, 
normalized by $\Delta_0$, at the same values of $I$.
The inset correlates with the main plot, as it shows 
that the singlet superconducting
correlations in S increase with $\alpha$ and are larger at $I=1$ than 
at $I=0.5$. 
The half-metallic case of $I=1$ illustrates that by having a single spin band 
populated at the Fermi surface,
Andreev reflection is suppressed, in effect keeping the superconductivity more 
contained within S.

Thus, 
we have shown that in clean FSF trilayers 
induced odd triplet
correlations,  
with $m=0$ and $m=\pm 1$ projections of the total spin, exist.
We have used a microscopic self-consistent
method to study the time and angular dependence of these triplet correlations.
The
correlations in all 3 triplet channels 
were found, at times $\tau\equiv \omega_Dt \gtrsim \tau_c$, where 
$\tau_c \approx 4$,  to be long ranged 
in both the F and S regions.
Finally, study of
the condensation energy revealed that the ground state energy is always lowest
for antiparallel magnetizations.


\acknowledgments
This project was
supported in part by a grant of HPC resources from 
the ARSC  at the University of Alaska Fairbanks 
(part of the DoD HPCM program)  and by the University of Minnesota
Graduate School.

\end{document}